\documentclass{article}
\usepackage[T1]{fontenc} 
\usepackage[utf8]{inputenc} 
\usepackage{ismir,amsmath,cite,url}
\usepackage{graphicx}
\usepackage{color}
\usepackage{amssymb}
\usepackage{booktabs}
\usepackage{textcomp}
\usepackage{subcaption}

\usepackage{todonotes}
\usepackage{siunitx}
\usepackage[bookmarks=false,hidelinks]{hyperref}
\urldef{\ourgithub}\url{https://github.com/fdlm/listening-moods}

\title{Mood Classification Using Listening Data}

\multauthor
{
	Filip Korzeniowski$^1$  \hspace{1cm} 
	Oriol Nieto$^1$         \hspace{1cm}
	Matthew C. McCallum$^1$
} 
{\bfseries{
	Minz Won$^2$            \hspace{1cm}
	Sergio Oramas$^1$       \hspace{1cm}
	Erik M. Schmidt$^3$
}\\
	$^1$ Pandora Media LLC., Oakland, California, USA\\
	$^2$ Music Technology Group, Universitat Pompeu Fabra, Barcelona, Spain\\
	$^3$ Netflix Inc., Los Gatos, California, USA
}
\def\authorname{Filip Korzeniowski, Oriol Nieto, Matthew C. McCallum, Minz Won, Sergio Oramas, Erik M. Schmidt}

\newcommand{\imgscale}{1.}
\begin{document}

\newcommand{\PandoraSet}{\mbox{AllMusic}}
\newcommand{\TP}{\mbox{TP-L}}
\newcommand{\PWF}{\mbox{P-L}}
\newcommand{\MCNMSD}{\mbox{MCN-MSD-A}}
\newcommand{\MCNMTT}{\mbox{MCN-MTT-A}}
\newcommand{\SCC}{\mbox{SCC-A}}
\newcommand{\numdata}{\num{66993}}
\newcommand{\numdatashort}{67k}

\maketitle
\begin{abstract}
The mood of a song is a highly relevant feature for exploration and recommendation in large collections of music. These collections tend to require automatic methods for predicting such moods. In this work, we show that listening-based features outperform content-based ones when classifying moods: embeddings obtained through matrix factorization of listening data appear to be more informative of a track mood than embeddings based on its audio content. To demonstrate this, we compile a subset of the Million Song Dataset, totaling \numdatashort{} tracks, with expert annotations of 188 different moods collected from \PandoraSet{}. Our results on this novel dataset not only expose the limitations of current audio-based models, but also aim to foster further reproducible research on this timely topic.
\end{abstract}
\section{Introduction}\label{sec:introduction}

The estimation of moods that a given music track might evoke or empathize with is a relevant task that has been active in the Music Informatics Research (MIR) community for years~\cite{kim_musicemotionrecognition_2010}.
This task, which is also known as music emotion recognition, has become even more prominent thanks to the advent of streaming music services with massive collections, where understanding the set of moods of each of their tracks could strongly impact the navigation, discovery, and recommendations of such collections~\cite{schedl_currentchallengesvisions_2018}.
This task has been typically approached in two different ways:
i) regressing a continuous mood space such as the Arousal-Valence one~\cite{russell_circumplexmodelaffect_1980}, and then clustering such space to obtain a specific mood vocabulary~\cite{soleymani_1000songsemotional_2013};
or ii) classifying a given track into one or more moods, thus becoming a multi-label classification problem with a fixed vocabulary~\cite{chowdhury_explainablemusicemotion_2019}, which can be seen as a sub-task of the broader audio tagging problem~\cite{pons_endtoendlearningmusic_2018}.
In this work, we focus exclusively on the second approach, since it can directly impact search-by-mood applications, while methods like metric learning can potentially overcome the limitation of the fixed vocabulary~\cite{choi_zeroshotlearningaudiobased_2019}.

Framed under the context of music recommendation, mood recognition is particularly interesting. 
It has been shown that listener personality correlates not only with musical taste~\cite{rentfrow_remieveryday_2003,Zangerle2018}, but also with genre~\cite{ferwerda_personalitytraitsmusic_2017}, which makes the development of psychologically inspired approaches one of the most compelling challenges for recommender systems~\cite{schedl_currentchallengesvisions_2018}.
Thus, several related techniques have been presented: FocusMusicRecommender~\cite{yakura_focusmusicrecommendersystemrecommending_2018} makes use of the listener's behavior history to play tracks that are appropriate given the current listener's level of concentration. By incorporating the Five Factor Model~\cite{costa_revisedneopersonality_1992}, collaborative filtering~\cite{koren_matrixfactorizationtechniques_2009} is enhanced with personality embeddings~\cite{fernandez-tobias_alleviatingnewuser_2016}.
Moreover, emotions from a microblogging service have been exploited to implement an emotion-aware recommendation system~\cite{deng_exploringuseremotion_2015}.
Such techniques employ data beyond the actual audio signal to enhance mood-based recommenders, inspiring us to make use of listening data to classify moods to potentially improve the navigation and recommendation of large music catalogs.

The contribution of this work to the task of mood prediction is two-fold:
i) we assemble a set of \numdatashort{} tracks from the Taste Profile subset from the Million Song Dataset (MSD)~\cite{bertin-mahieux_millionsongdataset_2011} and match them with human-annotated moods available from \mbox{AllMusic}.\footnote{\url{https://www.allmusic.com/}}
This is, to the best of our knowledge, the largest expert-annotated mood dataset available.
And ii) by running several experiments on this proposed dataset we show how listener data are much more accurate at classifying moods than current audio-based approaches. 
Similarly to~\cite{hu_whenlyricsoutperform_2010}, where its authors discuss how lyrics can be useful to predict moods better than actual audio, and following the music recommendation approaches described above, we further argue that listening embeddings yield superior results due to their ability to capture information that is not straightforward to be extracted from pure audio content only.

The rest of the article is structured as follows:
in Section~\ref{sec:method} we give a formal definition of the mood classification problem.
In Section~\ref{sec:mooddata} the data employed in this work are described.
We then detail the mood classification experiments in Section~\ref{sec:exp}.
The results of these experiments are discussed in Section~\ref{sec:res}.
Finally, we draw conclusions and consider potential future directions in Section~\ref{sec:conclusion}.

\section{Mood Classification}\label{sec:method}

Predicting moods evoked by music is often treated as an audio classification problem in the MIR community,\footnote{\url{https://www.music-ir.org/mirex/wiki/2019:Audio_Classification_(Train/Test)_Tasks}} where audio data are almost exclusively used as input.
In this section we give an overview of this task and its current approaches.

\subsection{Problem definition}\label{sec:problem_definition}

Mood tagging is a multi-label classification problem, and can be considered a subset of the broader audio tagging task where only those tags that represent moods are considered.
Formally, let $\mathbf{x} \in \mathbb{R}^E$ be an embedding representing a given track, where $E$ is the number of dimensions in the embedding. Each track is associated with a set of mood tags from a mood vocabulary $\mathcal{T}$ (e.g., ``energetic,'' ``gloomy,'' ``happy''), represented by a binary indicator vector $\mathbf{y} \in \{0, 1\}^{|\mathcal{T}|}$. We aim at predicting the set of mood tags associated with the track, using a learnable function $f$ that computes the predicted label vector $\hat{\mathbf{y}} = f(\mathbf{x})$.

Note that $\mathbf{x}$ can be extracted from any source of data representing the track. In our case, we will use audio- and listening-based embeddings.

Other approaches have also framed emotion prediction as a regression problem of an $n$-dimensional continuous space~\cite{soleymani_1000songsemotional_2013}, where the 2D Arousal-Valence model~\cite{russell_circumplexmodelaffect_1980} is the most widely used.
While this approach has the benefit of considering moods that are not constrained by a specific vocabulary, in this work we focus on the multi-label classification approach due to the direct application to potential user-based scenarios such as search by typing or by voice.

\subsection{Current Approaches}
The current state of the art largely approaches music mood prediction via audio analysis. Early approaches identified spectral contrast as an informative representation \cite{jiang_musictypeclassification_2002}, and a number of other authors confirmed this finding as well as a variety of other standard audio features \cite{kim_musicemotionrecognition_2010, schmidt_modelingmusicalemotion_2011, wang_modelingaffectivecontent_2015, schedl_interrelationlistenercharacteristics_2018}. While the relationship between mood and spectral representations remains non-obvious, previous work has shown that human subjects annotate reconstructions 
from these representations with reasonable consistency to their original form \cite{schmidt_relatingperceptualfeature_2012}. 
Still, the problem remained far from solved.

In moving towards increasing model complexity, most approaches have incorporated deep learning methods that seek to learn their own representations~\cite{schmidt_learningrhythmmelody_2013}.
In addition to prediction, audio-based approaches have also been extended to the problem of segmentation~\cite{aljanaki_emotionbasedsegmentation_2015}. More recent approaches have expanded to multi-modal representations by combining lyrics~\cite{delbouys_musicmooddetection_2018} and others have focused on interpretability of these complex models~\cite{chowdhury_explainablemusicemotion_2019}. At the time of writing, the authors are not aware of any models which leverage features derived from user interactions to estimate the moods of a music track. 

\section{Data}\label{sec:mooddata}

The data we collected for this work are derived from various sources: \PandoraSet{} 
provides mood annotations; The Echo Nest Taste Profile~\cite{mcfee_millionsongdataset_2012}, mapped to tracks in the Million Song Dataset, adds listening data; finally, 7-digital contributes 30s previews as audio data.

We link \PandoraSet{} data to the MSD by fuzzy string matching of artist and track names, and requiring track lengths to be within $\pm$10s. This results in a dataset of \numdata{} matched tracks in total, which we call the AllMusic Mood Subset (AMS). As opposed to other music tagging datasets, such as the LastFM Set~\cite{hu_improvingmoodclassification_2010,hu_whenlyricsoutperform_2010,cano_musicmooddataset_2017}, AMS provides a large vocabulary of mood tags annotated by music experts. While the AllMusic annotations are proprietary, they can be freely consulted on their website and, moreover, are available to be licensed.

Finally, we randomly split the AMS into 80\% training, 10\% validation, and 10\% test, resulting in \num{53585}, \num{6695}, \num{6713} tracks respectively. The splits are available online\footnote{\ourgithub} to ensure comparability of future results.

\subsection{Mood Data}\label{subsec:mooddata}

\begin{figure}[t]
 \centerline{
 \includegraphics[scale=\imgscale]{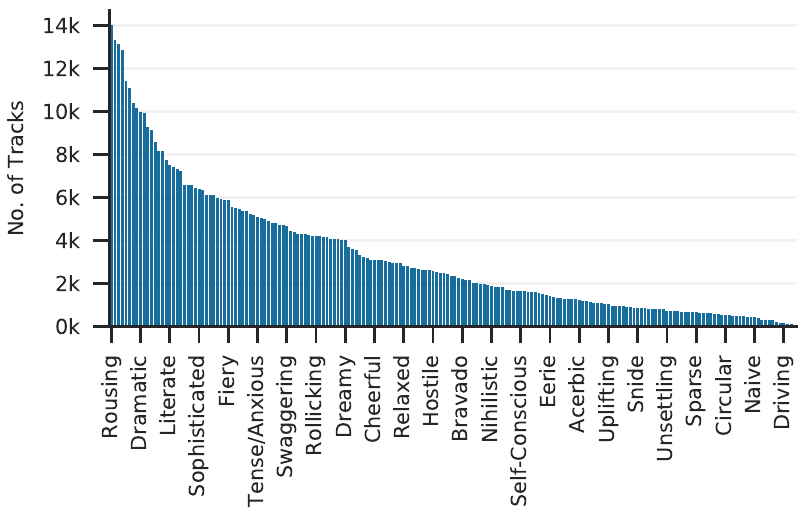}}
 \caption{Number of tracks per annotated mood in the AMS. Due to space limitations, only the names of a subset of mood tags are shown.}
 \label{fig:trackspertag}
\end{figure}

\begin{figure}[t]
 \centerline{
 \includegraphics[scale=\imgscale]{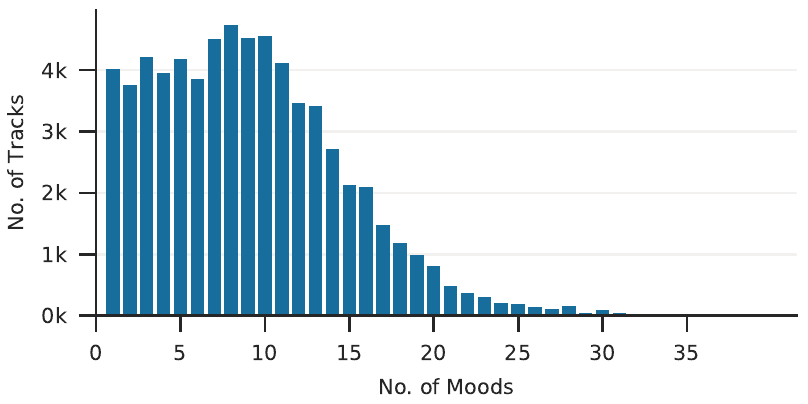}}
 \caption{Number of moods annotated per tracks in AMS.}
 \label{fig:tagspertrack}
\end{figure}

\begin{table}[]
\centering
\begin{tabular}{lrlr}
\toprule
\textbf{Top}  & \textbf{Count}   & \textbf{Bottom} & \textbf{Count} \\ 
\midrule
Rousing     & \num{14018}    & Melodic       & \num{95}        \\
Reflective  & \num{13330}    & Animated      & \num{140}       \\
Energetic   & \num{13153}    & Powerful      & \num{148}       \\
Earnest     & \num{12873}    & Driving       & \num{163}       \\
Passionate  & \num{11438}    & Introspective & \num{176}       \\ 
Confident   & \num{11092}    & Flowing       & \num{218}       \\ 
Amiable     & \num{10424}    & Positive      & \num{307}       \\ 
Intimate    & \num{10188}    & Stately       & \num{310}       \\ 
Dramatic    & \num{10014}    & Giddy.        & \num{315}       \\ 
Playful     & \num{9952}     & Thoughtful    & \num{340}       \\ 
\bottomrule
\end{tabular}
\caption{10 top and bottom mood tags based on the number of tracks they have been annotated in the AMS.}
\label{tab:trackspertag}
\end{table}

The mood information that we employ in this work has been human-annotated by experts from AllMusic.
These data were previously employed for mood classification~\cite{hu_exploringmoodmetadata_2007, bischoff_musicmoodtheme_2009} and lyrics sentiment detection~\cite{malheiro_classificationregressionmusic_2016}.
The mood tags are annotated at an album level, and we unfold them such that each track is assigned its album-level moods.

The total number of mood tags available is 188. As previous work noted~\cite{hu_exploringmoodmetadata_2007}, many tags may describe similar moods (such as ``Romantic'' and ``Sensual''), which tend to co-occur, and can be clustered into a smaller number of groups. While we can confirm this by performing manual and/or data-driven explorations on the co-occurrence matrix,
we intentionally kept the original annotations. For one, we expect modern machine learning methods to cope with large and possibly overlapping vocabularies. For another, these tags were curated by expert annotators to specifically describe how music feels; while they might characterize similar concepts, they could also provide a more nuanced view of a song's mood.

To give a better notion of the moods in this dataset, in Figure~\ref{fig:trackspertag} we depict the histogram of number of tracks per mood tag, which follows a typical long-tail distribution.
The 10 top and bottom annotated mood tags can be seen in Table~\ref{tab:trackspertag}.
As we can see, ``Rousing'' is the most frequent mood, which appears in \num{14018} tracks.
On the other hand, ``Melodic'' is the least frequent one, associated with only 95 tracks.
On average across the dataset, there are \num{3258.6}$\pm$\num{2961.3} tracks for each tag, with a median of \num{2385}.
Furthermore, Figure~\ref{fig:tagspertrack} shows the distribution of number of mood tags per track. It can be seen that most tracks have 13 moods or less, with an average of 9.1$\pm$5.7 tags per track and the median centered at 9.

\subsection{Audio Data}\label{subsec:audiodata}

Since the AMS is a subset of the MSD, we gather the audio data by obtaining the 7-digital 30 second previews associated with all MSD tracks.
These are 128kbps mp3 stereo files sampled at 44.1kHz.

\subsection{Listening Data}\label{subsec:listening}

We make use of the Taste Profile from the MSD to obtain listening data.
These data contain over $28$ million play counts from undisclosed partners associated with $L=\num{1019318}$ listeners and $S=\num{384546}$ tracks.

\begin{figure}[t]
 \centerline{
 \includegraphics[scale=\imgscale]{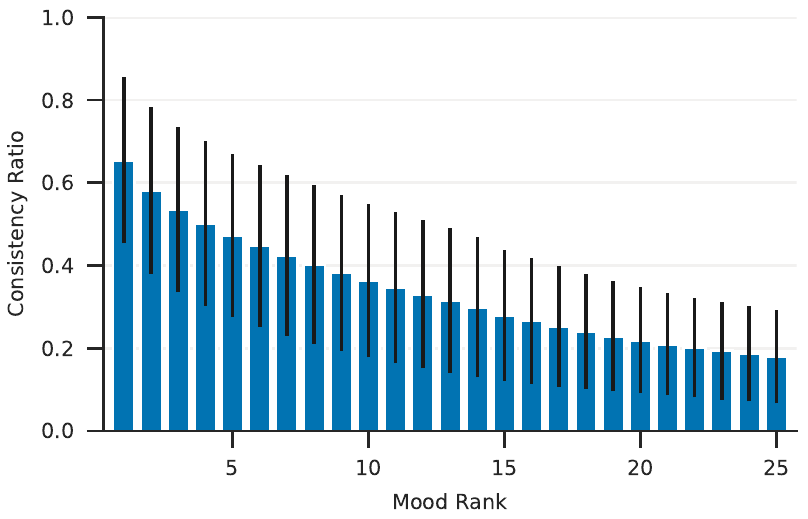}}
 \caption{Consistency ratios for the top 25 most popular moods for each user in the AMS.}
 \label{fig:consistency}
\end{figure}

We motivate the usage of such data in the context of mood classification by showing the relationship between listening habits and the moods of the tracks played, thus arguing that such embeddings are likely to contain relevant data when predicting moods.
By mapping the tracks in this set with the moods from the AMS (and thus reducing the set of listeners down to \num{1012825}, the track set down to \numdata{}, and the play counts down to \(\sim\)9 million) we observe that listeners play music that tends to be consistent in terms of its mood.
We define the consistency ratio of a mood as the fraction of times it appears in the listening history of a given user.
Figure~\ref{fig:consistency} shows the consistency ratio of the n\textsuperscript{th} most popular moods aggregated across users.
More specifically, 65.4\% of all plays by a given user contain the most popular mood tag for that user; similarly, around 50.1\% of a user's plays are annotated with their 4\textsuperscript{th} most popular mood; etc.
This exhibits the potential benefits of using listening data, as we confirm in the results of our experiments described next.

\section{Experiments}\label{sec:exp}

As described in Section~\ref{sec:problem_definition}, we treat mood prediction as a multi-label classification problem, with a function $f$ predicting mood tags $\hat{\mathbf{y}}$ from an input embedding $\mathbf{x}$. The input embedding can stem from different sources, such as listening- or audio-based features; we will refer to this as the \emph{embedding type}. We will use mostly open models trained on publicly available datasets in this work. As we will see, our conclusions follow from these results alone. Furthermore, we will show results for proprietary models trained on in-house listener feedback data. While we acknowledge that these additional results are hardly reproducible without access to our data and methods, they demonstrate how our findings translate to an industrial scale, and are thus a meaningful addition to this work.

\subsection{Evaluation Metrics}

Our goal is to compare the predictive performance of each embedding type, \emph{i.e.}, given an input embedding of a certain type, how well the predicted moods $\hat{\mathbf{y}}$ resemble the true moods $\mathbf{y}$ associated with a track. To quantify this, we will use macro-averaged \emph{average precision} as the main evaluation metric, as is commonly used in multi-label classification. Average precision summarizes the precision-recall curve in a single number, and is defined as
\begin{equation}
AP = \sum_n (R_n - R_{n-1})\cdot P_n,
\end{equation}
where $R_n$ and $P_n$ are the recall and precision at the $n$\textsuperscript{th} threshold at which the recall changes. Note that we use \emph{macro averaging}---we first compute AP for each mood tag, and then average them to calculate the final result.

In our mood prediction setup, there are two main questions we need to consider: how do we arrive at the input embedding $\mathbf{x}$, and how we model and train $f$. Let us first explore the various embedding types, before we take a detailed look at $f$. 

\subsection{Audio-Based Models}

Current mood prediction systems typically use audio-based features as input. In this work, we use several audio models, pre-trained on different datasets with varying sizes. This ensures that our results are not specific to a type of model. 

\subsubsection{Musicnn}

We employ Musicnn~\cite{pons_musicnnpretrainedconvolutional_2019}---a spectrogram-based convolutional neural network (CNN) for audio tagging---as the main pre-trained audio-based baseline. It is openly available\footnote{\url{https://github.com/jordipons/musicnn}} and achieves state-of-the-art results. We compare two variants of this model: a smaller one, trained on \(\sim\)19k tracks from the MagnaTagATune dataset~\cite{law_evaluationalgorithmsusing_2009}, which we will refer to as \emph{\MCNMTT{}}; and a larger one, trained on \(\sim\)200k tracks from the Million Song dataset, which we will name \emph{\MCNMSD{}}. Both variants come pre-trained to predict 50 tags, a subset of which can be associated with moods. We refer to Musicnn's documentation for further details on its training scheme.

Musicnn is trained to predict tags for 3-second snippets of audio; however, our setup requires a single \emph{embedding per track}. Thus, instead of the final output, we extract the activation of the penultimate layer of the model as embedding. We first compute embeddings of consecutive non-overlapping audio snippets of 3 seconds, and then average all snippet embeddings to form the track-level embedding. This results in a 200-dimensional vector for \MCNMTT{}, and a 500-dimensional vector for \MCNMSD{}. Such global averaging operations are common for music tagging~\cite{pons_endtoendlearningmusic_2018}.

\subsubsection{Short-Chunk CNN}

We train a short-chunk CNN~\cite{minzwon_evaluationcnnbasedautomatic_2020} from scratch on the 54k training tracks in the AMS. This simple but powerful model feeds a Mel-spectrogram through a 7-layer CNN with 3$\times$3 filters, 2$\times$2 max-pooling layers, and a fully connected layer before the output. For a detailed look into the training regime and architecture, we refer to the original paper.

Since this model was trained directly for mood prediction on the AMS, there is no need for transfer learning as described in  Section~\ref{sec:transfer_learning}. This is a double-edged sword: although the model is focused on the task at hand, it has to learn a large vocabulary of tags from the limited data provided by our dataset. We will refer to this model as \emph{\SCC{}}

\subsection{Listening-Based Models}

In contrast to audio-based models, listening-based ones consider user-song interaction as source data. This \emph{listening data} comes in the form of a sparse feedback matrix $Y \in \mathbb{N}^{L \times S}$, where $y_{l,s}$ is a cell in $Y$ representing the number of times the listener $l$ has either played or rated the song $s$. The former is called \emph{implicit} feedback, while the latter is referred to as \emph{explicit} feedback. Factorizing $Y$ using factorization rank $E$ (corresponding to the desired embedding dimensionality) yields dense track embeddings $\mathbf{x} \in \mathbb{R}^E$: the input to our mood prediction model.

\subsubsection{Taste-Profile Factorization}

We use listening data from the complete Taste Profile of 28M play counts to obtain song embeddings by applying weighted matrix factorization using alternating least squares~\cite{hu_collaborativefilteringimplicit_2008} with a rank of $E=200$ (chosen empirically). 
These data contain relevant information about the track defined exclusively with implicit feedback: how many times which listeners have listened to which songs. We will call these embeddings \emph{\TP{}}.

\subsubsection{Proprietary Factorization}\label{sec:p_wmf}

Large music streaming services possess much larger and more detailed listening data than openly available resources. To see how the results on open datasets translate to industrial settings, we derive 200-dimensional embeddings from more than 100B in-house explicit user ratings over the whole music catalog, by applying a weighted matrix factorization algorithm. These embeddings will be referred to as \emph{\PWF{}}.

\subsection{Transfer Learning}\label{sec:transfer_learning}

Having computed track-level embeddings $\mathbf{x}$ from various sources, we need to map them to mood tags using a learnable function $f$. This is a transfer-learning scenario: the input embeddings are obtained from a model trained to solve a different (but related) task, such as collaborative filtering or general audio tagging, and then applied for mood prediction by learning $f$. 

We model $f$ as a multi-layer perceptron (MLP) with a binary indicator vector as output, such that $\mathbf{\hat{y}} = f\left(\mathbf{x}\right)$, where $\mathbf{\hat{y}} \in \left[0, 1\right]^{|\mathcal{T}|}$. Thresholding $\mathbf{\hat{y}}$ gives us the set of predicted moods. We train $f$ for each embedding type by minimizing the binary cross-entropy between predicted vectors $\mathbf{\hat{y}}$ and target vectors $\mathbf{y}$ obtained from the true mood tags.

The performance of MLPs heavily depends on the choice of hyper-parameters. To enable a fair comparison, we optimized hyper-parameters for each embedding type individually using Bayesian optimization~\cite{snoek_practicalbayesianoptimization_2012}, monitoring average precision on the validation set. To limit the computational cost, we only used \TP{} and \MCNMSD{} as input embeddings, since they are the main points of comparison. Each setup enjoyed the same, fixed computational budget of 2 days on a single Tesla M40 GPU, which translates to around 200 trials per setup. Table~\ref{tab:hyper-parameters} shows details on the search space and the best found configurations. We see that for both embedding types, the best models reach the upper limit of our search space, which indicates that even larger models might lead to better results. However, we saw diminishing improvements for large models, so we do not expect much further improvement.

\begin{table}[]
\small
\centering
\begin{tabular}{@{}llrr@{}}
\toprule
\textbf{}                               & \textbf{Domain}   & \textbf{\TP{}} & \textbf{\MCNMSD{}} \\ \midrule
N\textsuperscript{\underline{o}} layers & $[2 .. 4]$        & 4           & 4                \\
N\textsuperscript{\underline{o}} units  & $[1500 .. 4000]$  & \num{3909}  & \num{3933}       \\
learning rate                           & $[0.0001, 0.005]$ & \num{4e-4}  & \num{5e-4}       \\
dropout~\cite{srivastava_dropoutsimpleway_2014}  & $[0, 0.5]$        & 0.25        & 0.25             \\
weight decay                            & $[0, 0.0001]$     & 0           & \num{1e-6}       \\ \bottomrule
\end{tabular}
\caption{Hyper-parameters optimized with Bayesian optimization, and best found configurations for each embedding type. Search ranges were defined based on limited initial experiments. For dropout and weight decay, we quantized the interval by 0.125 and \num{1e-06}, respectively.}
\label{tab:hyper-parameters}
\end{table}

We initialize the MLP weights using Kaiming's method~\cite{he_delvingdeeprectifiers_2015}, and use a rectifier 
activation function~\cite{glorot_deepsparserectifier_2011} after each layer (the output layer uses a sigmoid). The input is standardized using mean and standard deviation estimated on the training set. We then train $f$ for 100 epochs using a cosine-annealed learning rate~\cite{loshchilov_sgdrstochasticgradient_2017} (without restarts) and a 1-epoch warm-up phase. During training, we monitor average precision on the validation set to select the best performing model parameters. 
    
The code to reproduce these experiments is available online.\footnote{\ourgithub}

\section{Results}\label{sec:res}

\begin{figure}[t]
 \centerline{
 \includegraphics[scale=\imgscale]{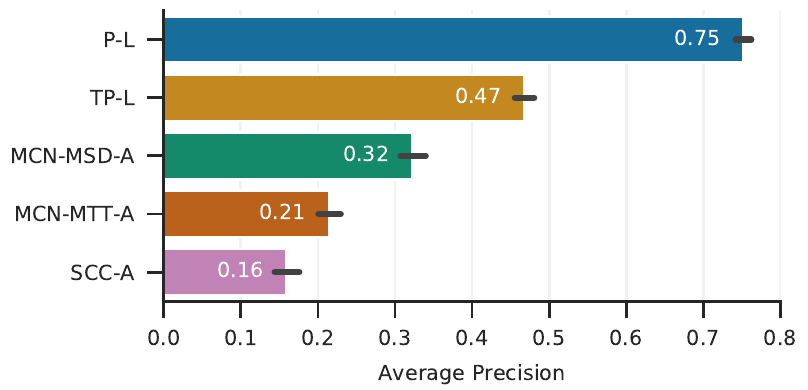}}
 \caption{Overall results for each model.}
 \label{fig:overall_results}
\end{figure}

Figure~\ref{fig:overall_results} shows the overall results of each embedding type. As mentioned before, our main analysis will be based on the results of open models on publicly available data. We will discuss the results of \PWF{} later.

We see that listening-based embeddings easily out-perform audio-based ones (\TP{} vs. \MCNMSD{}). We also see a variation within audio-based models. Our experiments were not designed to explain this variation, and the usual suspects offer insufficient clues: for example, dataset size might be an issue (200k for \MCNMSD{} vs. 19k for \MCNMTT{}), but \SCC{} was trained on the 54k training tracks from AMS with worse results---here, dataset size relative to vocabulary size might have been the issue. Further experiments, out of scope of this paper, are necessary to understand this in depth.
 
\subsection{Tag-Wise Results}

\begin{figure}[t]
\begin{subfigure}{\columnwidth}
  \centering
  \includegraphics[scale=\imgscale]{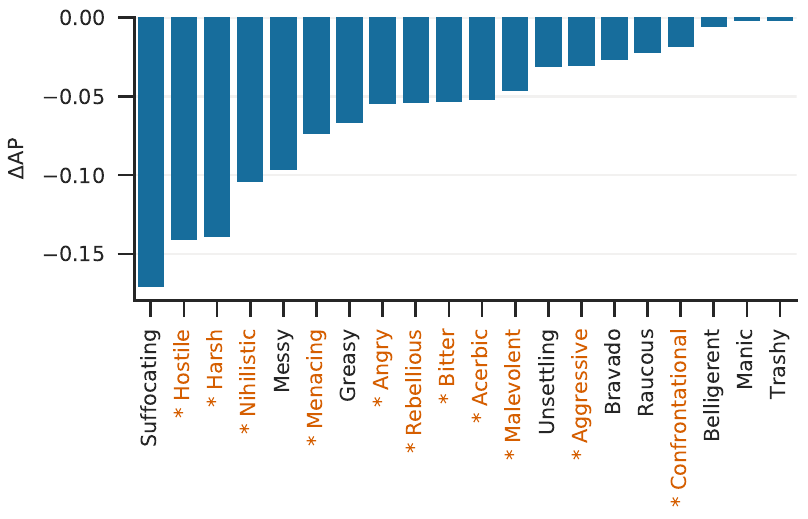}
  \caption{Tags favoring audio-based models.}
  \label{fig:audio_better}
\end{subfigure}
\\
\begin{subfigure}{\columnwidth}
  \centering
  \includegraphics[scale=\imgscale]{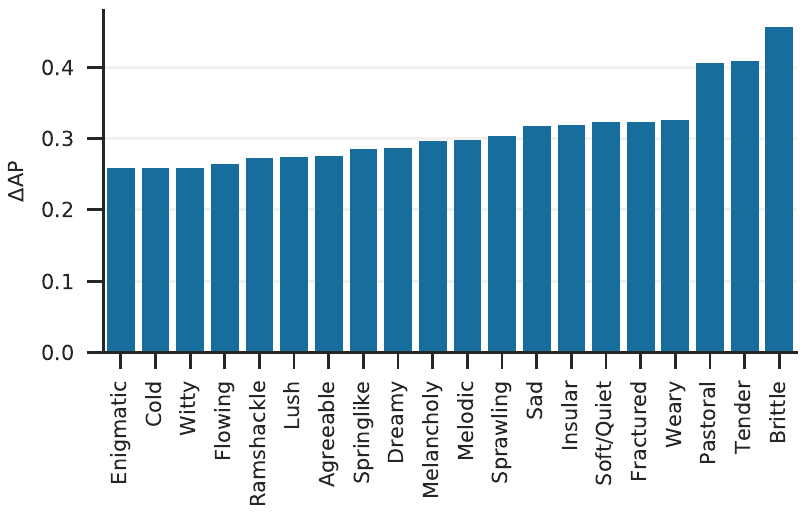}
  \caption{Tags favoring listening-based models.}
  \label{fig:cf_better}
\end{subfigure}
\caption{Difference of average precision between the best audio-based model (\MCNMSD{}), and the best
listening-based model on open data (\TP{}). Negative $\Delta$AP means the audio-based embedding performed better. The highlighted
tags in (\subref{fig:audio_better}) belong to the same mood cluster.
}
\label{fig:tag_wise_comparison}
\end{figure}

Even though the overall results are clear, some tags might be easier to predict from audio than from listening data. To explore this, we subtract the tag-wise average precision of \TP{} and \MCNMSD{}, and show the results in Figure~\ref{fig:tag_wise_comparison}. Indeed, we find 20 tags for which \MCNMSD{} out-performs \TP{}. Moreover, these tags seem to describe related moods. To verify this, we clustered the 188 moods using affinity propagation~\cite{frey_clusteringpassingmessages_2007}, resulting in 13 clusters. We see that 11 out of the 20 mood tags belong to the same cluster, as highlighted in Figure~\ref{fig:audio_better}. In contrast, the tags in Figure~\ref{fig:cf_better} come from a wider variety of clusters (not highlighted). This indicates that it is a single, coherent ``mood subspace'' on which audio data is better suited. 

\subsection{Results by Tag Frequency}

\begin{figure}[h]
 \centerline{
 \includegraphics[scale=\imgscale]{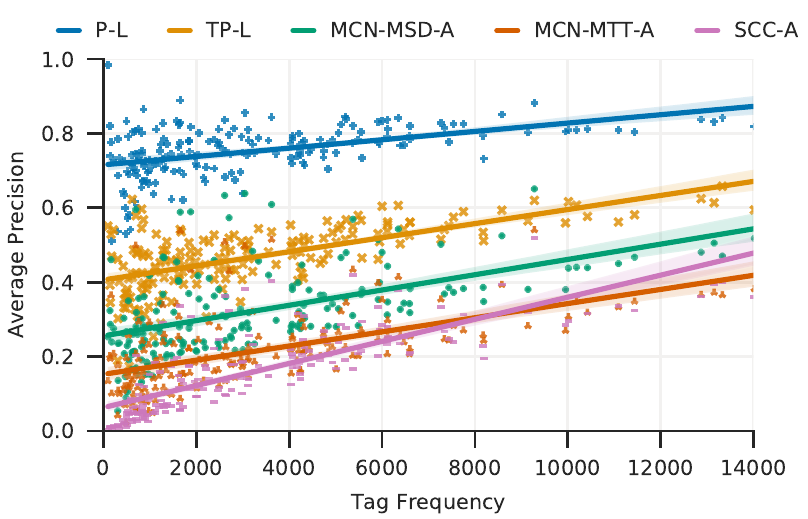}}
 \caption{Results per tag frequency. Dots represent the average precision obtained by a tag,
 that occurs at a frequency shown on the x axis. Lines represent linear regression models, with shades indicating 95\% confidence intervals.}
 \label{fig:results_per_freq}
\end{figure}

As shown earlier, mood tags in the AMS are unevenly distributed: the least popular tag counts only 95 annotations, while the most popular track 14k. It is reasonable to assume that uncommon tags are more difficult to predict than common ones. To evaluate this, we plot the average precision per tag depending on the tag frequency in Figure~\ref{fig:results_per_freq}. 

Although we see a direct relation between tag frequency and average precision, the extent is less than we expected. Furthermore, all embedding types seem to be equally affected: with the exception of \SCC{}, the regression slopes of both audio- and listening-based models are notably similar. The exception of \SCC{} indicates that tag sparsity may be an issue when training audio models from scratch, but not so when transfer-learning a model that has been trained on more balanced data. 

\subsection{Results of Proprietary Algorithms}

So far, we have discussed the results of open methods on publicly available datasets. However, the attentive reader has noticed that Figure~\ref{fig:overall_results} and \ref{fig:results_per_freq} demonstrate how \PWF{} performs even better than \TP{}. To explain the gap between \TP{} and \PWF{}, we can point to the different nature and amount of data they were trained on---28M implicit plays for the former, but more than 100B explicit ratings for the latter. The sheer amount of data (a factor of $\sim$3500) and the stronger signal provided by explicit feedback seem to be remarkably beneficial.

\subsection{Consistency of Audio-Based Models}

We have shown that listening-based models clearly out-perform audio-based models in mood prediction. To demonstrate this, we selected a wide variety of audio models that differed in multiple aspects: network architecture, training datasets, and training regime (pre-trained and trained from scratch).
Given these differences, we can ask if there are aspects of mood that current audio models are not capable to capture, but listening-based models can. We try to answer this question by exploring which embeddings capture similar mood information. If an embedding captures similar aspects of mood as another embedding, their tag-wise performance should be correlated---but not necessarily similar in magnitude, as one embedding might just perform better than the other.

\begin{figure}[t]
 \centerline{
 \includegraphics[scale=\imgscale]{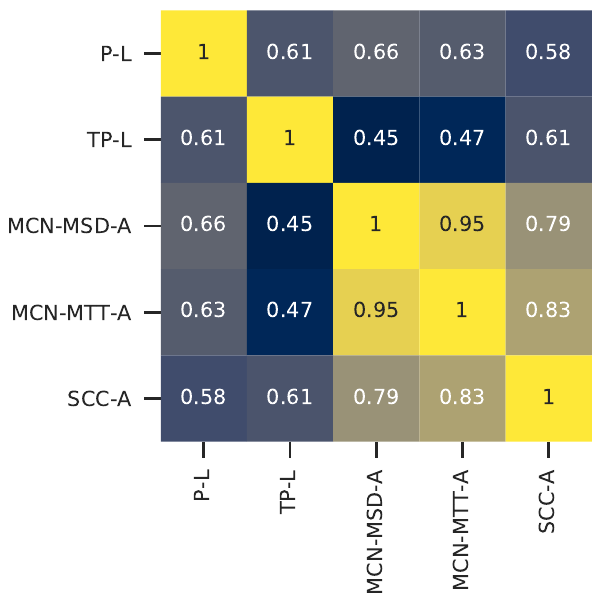}}
 \caption{Correlation between tag-wise results of different embeddings. We see that audio-based ones correlate strongly with each other, compared to weaker correlations between listening-based ones.}
 \label{fig:result_correlation}
\end{figure}

We show the correlation in tag-wise performance in Figure~\ref{fig:result_correlation}. The remarkable result is that regardless of their differences, the tag-wise results of all audio-based models are much more correlated than between audio- and listening-based embeddings. This indicates that audio-based models do capture similar aspects, even if they might not capture it equally well (as the difference between \MCNMSD{} and \SCC{} shows). This does not mean that the aspects current audio-based models are missing are not present in the audio at all---just that current models are not able to extract them. 

We do not observe a similar pattern for listening-based embeddings: \TP{} and \PWF{} show weaker correlation. At this time, we cannot provide a better explanation than referring to the different nature of explicit and implicit feedback data and the sizes of the two datasets. 

\section{Conclusions}\label{sec:conclusion}

In this work we have associated \numdata{} tracks from the Million Song Dataset with the \PandoraSet{} set to yield the AMS, the largest dataset available with the following data modalities: high quality human mood annotations, audio content, and listening data.
Furthermore, we have shown how listening data surpass audio-based embeddings when classifying moods in the proposed dataset.
The notable differences in performance between listening- and audio-based models suggest that either i) current state-of-the-art audio models are not capable of successfully extracting certain mood information about a given track; and/or ii) such mood information is not necessarily present in the audio content, and thus the usage of other signals such as listening information may be required to obtain more accurate results.
With these findings, we encourage researchers to employ data beyond audio content when estimating the mood of a track.
In the future, we look to further scrutinize the tags to better understand which moods might be more suitable to be extracted by which type of input representation.
Moreover, and along these lines, we would like to address this task in a multi-modal manner, combining different sources to potentially improve performance of this compelling and timely problem.

\newpage

\section{Acknowledgements}

The authors F. Korzeniowski and O. Nieto contributed equally to this work.

\bibliography{references}

\end{document}